# Tricritical Point and the Doping Dependence of the Order of the Ferromagnetic Phase Transition of $La_{1-x}Ca_xMnO_3$

D. KIM,[*] B. REVAZ, B. L. ZINK, AND F. HELLMAN

DEPARTMENT OF PHYSICS, UNIVERSITY OF CALIFORNIA, SAN DIEGO, LA JOLLA, CA 92093

J. J. RHYNE

DEPARTMENT OF PHYSICS, UNIVERSITY OF MISSOURI, COLUMBIA, MO 65211

J. F. MITCHELL

MATERIAL SCIENCE DIVISION, ARGONNE NATIONAL LABORATORY, ARGONNE, IL 60439

We report the doping dependence of the order of the ferromagnetic metal to paramagnetic insulator phase transition in $La_{1-x}Ca_xMnO_3$. At x = 0.33, magnetization and specific heat data show a first order transition, with an entropy change (2.3 J/molK) accounted for by both volume expansion and the discontinuity of M ~ 1.7 $\mu_B$ via the Clausius-Clapeyron equation. At x = 0.4, the data show a continuous transition with tricritical point exponents $\alpha$ = 0.48±0.06, $\beta$ = 0.25±0.03, $\gamma$ = 1.03±0.05, and $\delta$ = 5.0±0.8. This tricritical point separates first order (x<0.4) from second order (x>0.4) transitions.

The magnitude of colossal magnetoresistance (CMR) of mixed valence manganites near $T_C$ is inversely related to their $T_c$, and may be related to the order of the ferromagnetic phase transition (FPT) [1,2]. The highest $T_C$ manganite $La_{1-x}Sr_xMnO_3$ (LSMO) with $0.2 < x < 0.5$ has a conventional second order FPT and the magnetoresistance is the lowest in the manganite family [3,4,5,6]. Lower $T_C$ manganites like $Pr_{1-x}Ca_xMnO_3$ (PCMO) with $0.15 < x < 0.3$ show a first order FPT and much higher CMR [7]. The $T_C$ and magnetoresistance of $La_{1-x}Ca_xMnO_3$ (LCMO) is between that of LSMO and PCMO and the FPT for LCMO for $x\sim0.33$ has been variously reported as second order and more recently as first order [6,8,9,10,11,12,13,14]. The source of a first order transition in LCMO has been suggested to lie in strong electron-phonon coupling and/or in an intrinsic inhomogeneity in these magnetic oxides which gives rise to competing coexisting ground states [15,16,17,18]. The high temperature paramagnetic state has larger volume than the low temperature ferromagnetic metallic state, with a Jahn-Teller distortion to accommodate localized $e_g$ electrons, so that a first order FPT with a discontinuous change in volume and entropy indicates a discontinuous change in the number of localized electrons. In LCMO, most attention has focused on $x \sim 0.33$, but neither $T_c$ nor the magnetoresistance are strong functions of composition for $0.3<x<0.45$. A recent neutron scattering study however found a significant difference in the temperature dependence of the spin wave stiffness near $T_c$ for $x =0.33$ and $0.4$, suggestive of a change in the order of the transition [12].

We have performed high precision specific heat and magnetization measurements through the FPT on LCMO with $x = 0.4$ and $0.33$. Samples were prepared by solid state reaction. The resulting sintered pellets were single phase as judged by powder x-ray diffraction. Oxygen stoichiometry was determined by triplicate iodometric titration, yielding an oxygen content of 3.00(1) for each sample. After sintering, samples were ground into coarse powders, with particle size large enough that we were able to use individual particles for the measurements. Specific heat and magnetization measurements were made on samples taken from the same processing batch. For specific heat, x=0.33 and 0.4 samples were both 0.3 x 0.5 x 0.5 mm$^3$, and weighed 429 and 491 μg respectively. Use of sensitive SiN membrane-based microcalorimeters enables us to measure these small samples, reducing sample inhomogeneity. In addition to our standard high accuracy measurement technique, we also used a large ΔT method (related to the sweep method) to improve our temperature resolution near $T_c$. Further details of these measurements and the microcalorimeters may be found in refs. 5, 19 and 20. Measurements of magnetization M as a function of applied field $H_a$ and temperature T were made in a Quantum Design SQUID magnetometer. Samples were 0.7 x 0.8 x 1.2 mm$^3$ (3.75 mg) and 0.8 x 0.8 x 0.7 mm$^3$ (2.65 mg) for x=0.33 and 0.4 respectively.

For a second order FPT near the critical temperature, where a correlation length is defined, the specific heat $C_P$, spontaneous magnetization ($M_s \equiv M(H=0)$) and initial magnetic susceptibility ($\chi \equiv \partial M/\partial H|_{H=0}$) show power law dependence on the reduced temperature, $t \equiv (T - T_C) / T_C$ with critical exponents α, β, and γ, and at $T_C$ $M(H) \sim D\, H^{1/\delta}$. The critical exponents of a FPT depend on the type of ordering and dimensionality, but, as discussed by Huang, at a tricritical point, they are universal: α = 0.5, β = 0.25, γ = 1, and δ = 5.[21] Critical exponents are not defined at a first order transition. For a first order FPT, the magnetic field can shift the transition leading to a field dependent phase boundary $T_C(H)$.

Figure 1 shows magnetization data for x=0.4. The internal field H = $H_a$ - 4πDM is derived from the applied field $H_a$ by calculating the demagnetization correction from the low field $M(H_a)$ data. The uncorrected data is in ref. [20]. The M(H) data for temperatures near $T_c$ (256 to 270K) are plotted in Fig. 1a as a modified Arrot plot ($M^{1/\beta}$ vs $(H/M)^{1/\gamma}$), which is based on the Arrot-Noakes equation of state $(H/M)^{1/\gamma} = (T-T_c)/T_c + (M/M_s)^{1/\beta}$ [22]. The plot uses the optimal values of $\beta$ and $\gamma$ obtained by iterative approximation, discussed below. Assuming asymptotic critical behavior $(\chi)^{-1} = (H/M) \sim |t|^\gamma$ and $M_s \sim |t|^\beta$, a plot of $(H/M)^{1/\gamma}$ vs $M^{1/\beta}$ is linear, and isothermal curves become parallel straight lines (mean field values $\beta = 0.5$, $\gamma = 1$ give the regular Arrot plot $M^2$ vs $H/M$). The X- and Y-axis intercepts of each isothermal line yield $1/\chi^{1/\gamma}$ and $M_s^{1/\beta}$ for each T. The resulting $M_s$ vs T, log $M_s$ vs log t, $1/\chi$ vs T, and log $1/\chi$ vs log t plots are shown in Figs. 1b and c. The slopes of the log-log plots yield values for $\beta$ and $\gamma$, which are used to create a new modified Arrot plot and iterated until consistent. For the x = 0.4 sample, this procedure yields critical exponents $\beta = 0.25 \pm 0.03$ and $\gamma = 1.03 \pm 0.05$, the values used for Figs. 1a-c. Error bars come from the least squares fit analysis.

The critical exponent $\delta$ is found from log M vs log H at $T_c$. The critical isotherm $T_C$ = 265.5K lies between isothermal lines for 265 and 266K, which yield $\delta = 5.8$ and 4.2 respectively, as shown in Fig. 1d. Interpolating, we obtain $\delta = 5.0 \pm 0.8$.

Figure 2 shows the specific heat for the x=0.4 sample. In order to get the critical exponent $\alpha$, it is necessary to determine the non-critical background, which contains both magnetic and non-magnetic contributions. We used the analytical method of Kornblit and Ahlers [23], with fitting functions $C_P^\pm(t) = (A_\pm/\alpha)|t|^{-\alpha} + B + Ct$ where (+) corresponds to t > 0, (-) corresponds to t < 0, and B+Ct is the background. The analysis [described further in ref. 20] involves choosing a set of possible values for B and C. The best fit values B = 115.8 and C = 18.46 are obtained by a least squares minimization procedure (called a grid search method) [24]. The inset of Fig. 2 is a log-log plot of the resulting singular specific heat $C_S(t) = C_P(t) - (B+Ct) = (A/\alpha)|t|^{-\alpha}$. The data fit power law behavior to quite small t, with deviations (due possibly to sample inhomogeneity)

seen only below $t \sim 5*10^{-3}$ ($T - T_c \sim 1.3$ K). We find $\alpha = 0.48 \pm 0.06$; the uncertainty is determined from the two standard deviation contours in the $\chi^2(B,C)$ plot.

Turning now to x=0.33, we find extremely different behavior near $T_c$. Figure 3a shows M vs H for T between 250 and 290 K. For $T < T_c \sim 260$ K, the data are typical of a ferromagnet with small coercivity < 50 Oe and small hysteresis; the curvature in M(H) below 2000 Oe below 260K is due to domain reorientation. Just above $T_c$ however, we see S-shaped M(H) curves, indicative of a metamagnetic phase transition. Figure 3b shows an expanded plot for 264 K. For H < 2000 Oe, M(H) is linear and shows no hysteresis. Above 2000 Oe, there is hysteresis and positive curvature $d^2M/dH^2$, indicative of re-entrant ferromagnetism. In Fig. 3a, the onset of reentrant ferromagnetism for each T is marked by thick line segments. The resulting phase diagram is shown in Fig. 3c, which shows the slope of the coexistence line $dH/dT_C = 540$ Oe/K.

The specific heat of the x=0.33 sample (shown in Fig. 4) is also strikingly different from x=0.4. The peak at $T_c$ is much sharper, higher, narrower, and symmetric. Attempts to get a critical exponent $\alpha$ yield poor fits and values > 1, too large for any second order FPT model. To analyze the data, we made a third order polynomial background fit (shown in the figure) from $C_p$ data far from $T_C$, allowing for an offset at $T_c$. The inset shows $C_s = C_p$-background. Unlike three dimensional second order FPT, the shape of the peak is symmetric (note that the background determination does not affect the symmetric shape, as the peak is large compared to the background). The full width half maximum of the peak is 1.6 K (comparable to that seen in ref. 13); this is likely due to broadening of a delta function peak of the first order FPT, primarily from sample inhomogeneity as the experimental resolution is ~0.3K.

We use the magnetic extension of the Clausius-Clapeyron equation to analyze the PM/FM phase boundary shown in Fig. 3c. The differential Gibbs free energy dG = -SdT + VdP – MdH in both phases. Since $G_{FM} = G_{PM}$ for all points along the phase boundary line, $dG_{FM} = dG_{PM}$, which can be rewritten as $\Delta S = (S_{PM}-S_{FM}) = (V_{PM}-V_{FM})*dP/dT_C - (M_{PM}-M_{FM})*dH/dT_C = \Delta V/\{dT_C/dP\} - \Delta M/\{dT_C/dH\}$. In words, the entropy, volume, and magnetization changes across the phase boundary at a given pressure P, T, and H are related to the slopes $dT_C/dP$ and $dT_C/dH$ of the phase boundary surface which defines $T_C(P,H)$. The entropy change $\Delta S = 2.3$ J/molK is obtained by integrating $C_S/T$ through the phase transition. A 0.1% volume change through the FPT ($V_{PM}>V_{FM}$) of $La_{0.67}Ca_{0.33}MnO_3$ was reported by thermal expansion[25], yielding $\Delta V \sim 0.001*(35\times 10^{-6})$ m$^3$/mol (at low P). The pressure dependence of $T_C$ was variously reported as 1.6-2.2 K/kbar [25,26,27], yielding the entropy change due to volume expansion $\Delta V/\{dT_C/dP\} = 2.2$-1.6 J/molK. We take $dT_C/dH = 0.0019$ K/Oe from Fig. 3c and $\Delta M = M_{PM}|_{T_C^+} - M_{FM}|_{T_C^-}$ from Fig. 3a, which shows $M_s = 1.7$ $\mu_B$ at T = 258K, taking account of the low field domain reorientation, dropping to zero at 260K, giving $\Delta M|_{T_C}/\{dT_C/dH\} = -0.5$ J/mole K. Thus most of the entropy difference between the PM and FM states (~1.9±0.3 out of 2.3 J/mole K) comes from a volume change, with only ~0.5 J/mole K from magnetization.

It is thus clear that the FPT at x=0.33 is first order. It is also clear that for x = 0.4, the transition is continuous, with critical exponents $\alpha = 0.48$, $\beta = 0.25$, $\gamma = 1.03$, and $\delta = 5.0$, in excellent agreement with scaling exponent relations $\alpha+2\beta+\gamma = 2$ and $\gamma = \beta(\delta-1)$, and with the values for a tricritical point [21]. The existence of the x = 0.4 tricritical point sets a boundary between first order (x < 0.4) and second order (x > 0.4) FPTs within the ferromagnetic range (0.2 < x < 0.5). A change from first- to second-order FPT in the manganites with increasing carrier density x is found in several theories [15,16], in qualitative agreement with our data, but has not been previously experimentally observed.

In conclusion, we have performed specific heat and magnetization measurements on $La_{1-x}Ca_xMnO_3$ to study the nature of the FPT. At x = 0.33, the data show several signatures of a first order transition: a metamagnetic transition above $T_C$ due to reentrant ferromagnetism (including hysteresis), a discontinuous drop of magnetization at $T_C$ of 1.7 $\mu_B$, and an extremely narrow (1.6K FWHM) and symmetric specific heat peak. The change of entropy associated with the phase transition is $\Delta S \sim 2.3$ J/molK. The slope of the first order FPT line $dH/dT_C = 540$ Oe/K, determined from the re-entrant ferromagnetism. The Clausius-Clapeyron equation suggests the change of entropy comes primarily from the volume expansion (1.6 -2.2 J/molK), with a smaller contribution from the discontinuous change of magnetization at $T_c$. By contrast, for x = 0.4, the data show a continuous FPT with critical exponents $\alpha = 0.48 \pm 0.06$, $\beta = 0.25 \pm 0.03$, $\gamma = 1.03 \pm 0.05$, $\delta = 5.0 \pm 0.8$, in excellent agreement with tricritical point values. This tricritical point has not previously been observed and separates regions of first and second order FPT.

We would like to thank J. Lynn and D. Arovas for valuable discussions, and the NSF DMR for support (DK, BLZ, FH). This work was sponsored in part by the US Department of Energy Office of Science under Contract No. W-31-109-ENG-38.

**Figure Captions**

Fig. 1. a) Modified Arrot Plot of $La_{0.6}Ca_{0.4}MnO_3$; $\gamma = 1.03$, $\beta = 0.25$. Internal field H = applied field $H_a$ - $4\pi DM$ with D= 0.29 (nearly cubic) from low field M vs $H_a$. b) Spontaneous magnetization $M_s$ vs T, from Y-intercepts of a). Inset: $\log(M_S)$ vs $\log(|t|)$ with $T_C = 265.5K$ gives slope $\beta$. c) Inverse susceptibility $1/\chi$ vs T, from X-intercepts of a). Inset: $\log(1/\chi)$ vs $\log(t)$ with $T_C = 265.5K$ gives slope $\gamma$. (d) M vs H for T just above and below $T_c$. Inset: log M vs log H gives inverse slope $\delta = 5.0 \pm 0.8$.

Fig. 2. Specific heat of $La_{0.6}Ca_{0.4}MnO_3$ (symbols) with best fit (solid lines). Broken line is linear background (115.8 + 18 t) obtained by $\chi^2$ grid search method. Inset: $\log(C_S)$ vs $\log(|t|)$ with $C_S(t) = C_P(t) - (B+Ct) = (A/\alpha)|t|^{-\alpha}$, with best fit B and C; $\alpha = 0.5$ is the (negative) slope.

Fig. 3. a) M vs H of $La_{0.67}Ca_{0.33}MnO_3$ (for clarity, only decreasing field data are shown); demagnetization factor D=0.25 (nearly cubic) from low field M vs $H_a$. Thick black line segments show onsets of reentrant ferromagnetism. b) M vs $H_a$ at T = 264K, showing hysteresis above $|H_a| > 2000$ Oe. c) H-T phase diagram.

Fig. 4. Specific heat of $La_{0.67}Ca_{0.33}MnO_3$ (solid symbols) with polynomial background (open symbols) fit far from $T_c$. Inset: $C_P$ - background.

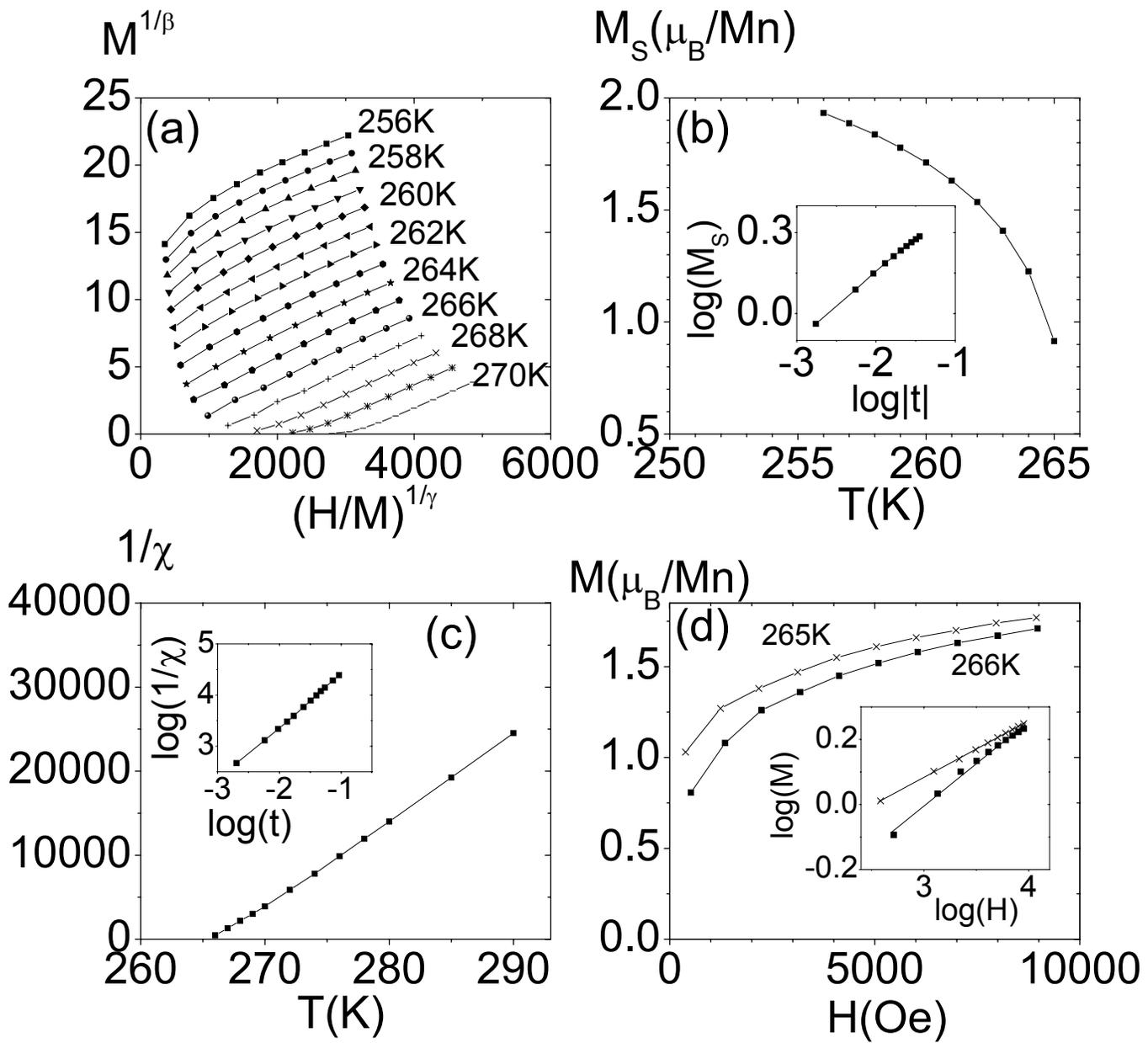

Fig.1

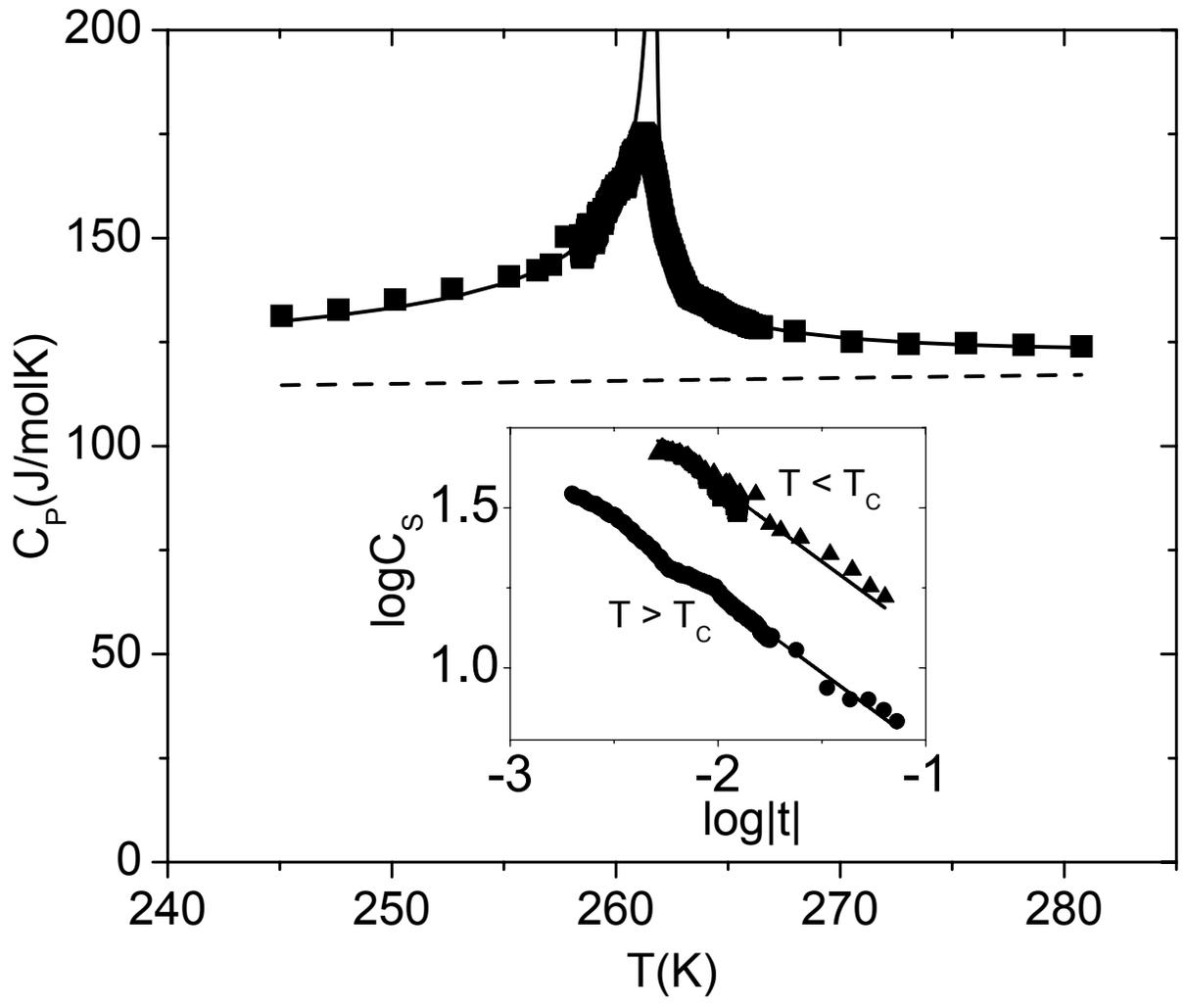

Fig.2

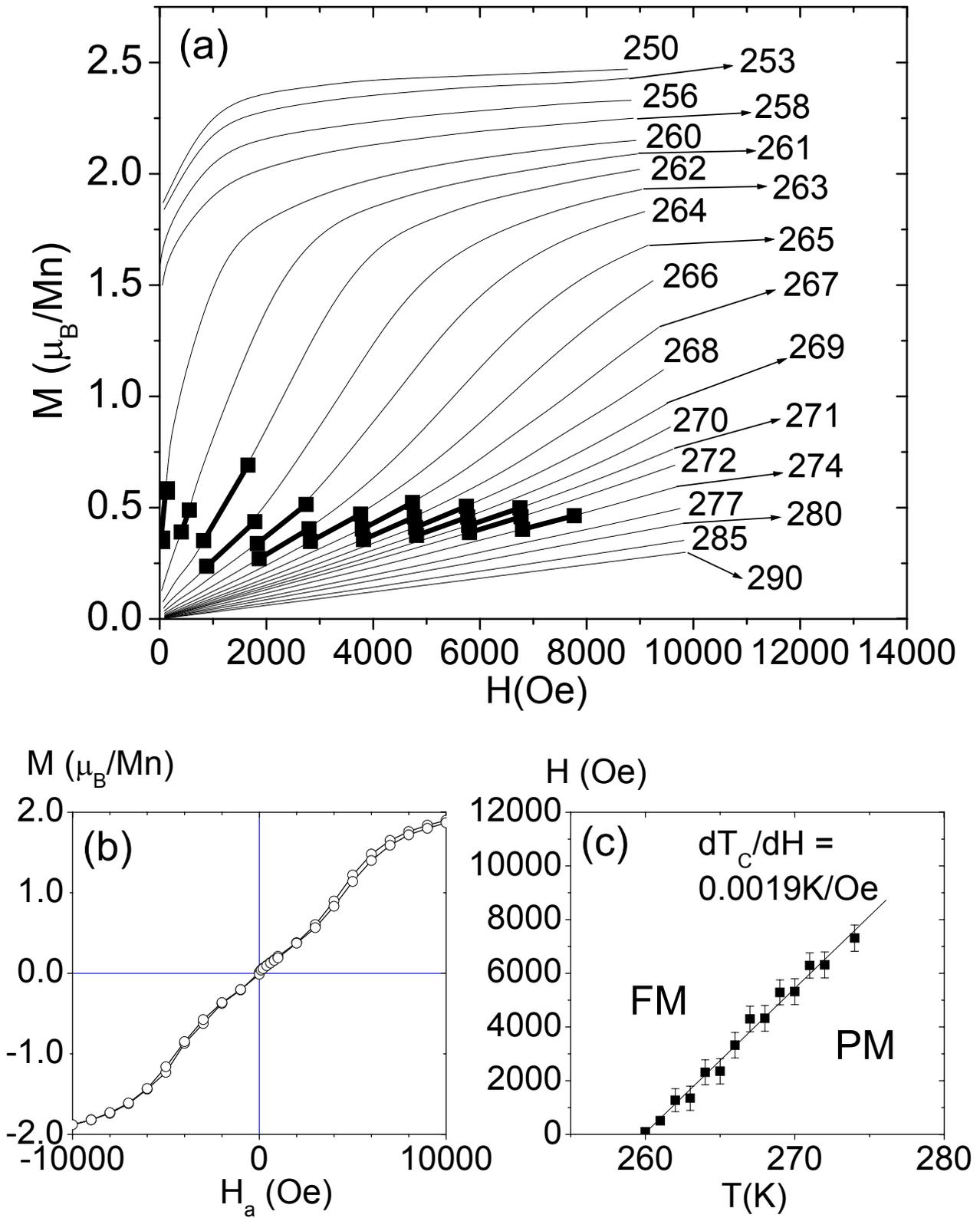

Fig.3

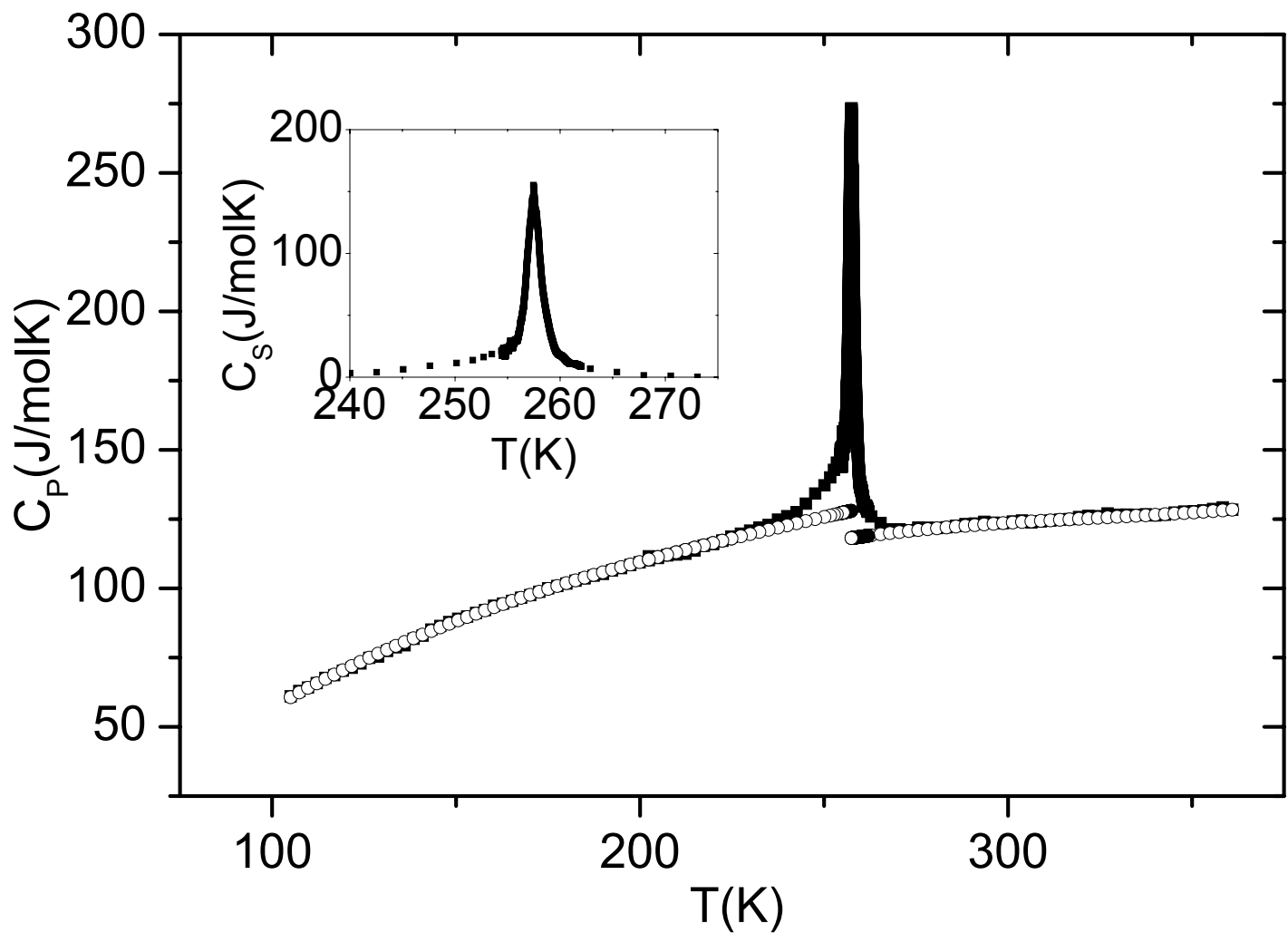

Fig.4